\documentstyle[11pt,aaspp4]{article} 

\newcommand{\ea}{et al.\ }
\newcommand{\subH}{_{HI}}
\newcommand{\uvunits}{ergs cm$^{-2}$ s$^{-1}$ Hz$^{-1}$ sr$^{-1}~$}   

\newcommand{\Lya}{Ly$\alpha$}

\newcommand{\lsim}{\lesssim}

\begin{document}

\title{THE METAGALACTIC IONIZING RADIATION FIELD   \\ 
    AT LOW REDSHIFT }  

\author{J. MICHAEL SHULL$^1$, DAVID ROBERTS$^2$, MARK L. GIROUX, \\
STEVEN V. PENTON, AND MARK A. FARDAL$^3$}

\affil{Center for Astrophysics and Space Astronomy, \\ 
Department of Astrophysical and Planetary Sciences,\\ 
University of Colorado, Campus Box 389, Boulder, CO 80309 \\
(mshull,giroux,spenton)@casa.colorado.edu \\
     \        \\
$^1$ also at JILA, University of Colorado and National
Institute of Standards and Technology \\
      \         \\
$^2$ Current address:  Physics Department, Cornell University,
Ithaca, NY 14853;
dcr12@cornell.edu \\
    \          \\
$^3$ Current address: Department of Physics \& Astronomy, University
of Massachusetts, Amherst, MA 01003;
fardal@weka.phast.umass.edu}

\begin{abstract}

We compute the ionizing radiation field at low redshift, 
arising from Seyferts, QSOs, and starburst galaxies.  
This calculation combines recent Seyfert luminosity functions,
extrapolated ultraviolet fluxes from our IUE-AGN database, and 
a new intergalactic opacity model based on {\it Hubble Space Telescope}
and {\it Keck} Ly$\alpha$ absorber surveys.  At $z = 0$ for AGN only, 
our best estimate for the specific intensity at 1 Ryd is 
$I_0 = 1.3^{+0.8}_{-0.5} \times 10^{-23}$ \uvunits,
independent of $H_0$, $\Omega_0$, and $\Lambda$. The one-sided ionizing 
photon flux is 
$\Phi_{\rm ion} \approx 3400^{+2100}_{-1300}$ photons cm$^{-2}$ s$^{-1}$,
and the H~I photoionization rate is $\Gamma_{\rm HI} =
3.2^{+2.0}_{-1.2} \times 10^{-14}$ s$^{-1}$, for $\alpha_s = 1.8$. 
We also derive $\Gamma_{\rm HI}$ for $z = 0 - 4$.
These error ranges reflect uncertainties in the spectral indexes for 
the ionizing EUV ($\alpha_s = 1.8 \pm 0.3$) and the optical/UV 
($\alpha_{\rm UV} = 0.86 \pm 0.05$), the IGM opacity model, the 
range of Seyfert luminosities ($0.001 L_\ast - 100 L_\ast$), and the 
completeness of the luminosity functions.
Our estimate is a factor of three lower than the most stringent 
upper limits on the ionizing background 
($\Phi_{\rm ion} < 10^4$ photons cm$^{-2}$ s$^{-1}$) 
obtained from H$\alpha$ observations in external clouds, and it lies
within the range implied by other indirect measures.  Starburst galaxies 
with a sufficiently large Lyman continuum escape fraction,
$\langle f_{\rm esc} \rangle \geq 0.05$, may provide a comparable background 
to AGN, $I_0(z=0) = 1.1^{+1.5}_{-0.7} \times 10^{-23}$ \uvunits.
An additional component of the ionizing background of this
magnitude would violate neither upper limits from H$\alpha$
observations nor the acceptable range from other measurements.

\end{abstract}


\newpage
\normalsize

\section*{1. INTRODUCTION }

The ionizing background that permeates intergalactic space
is of fundamental interest for interpreting QSO absorption
lines and interstellar high-latitude clouds.
Produced primarily by quasars, Seyfert galaxies, and other
active galactic nuclei (AGN), these Lyman-continuum (LyC) photons
photoionize the intergalactic medium (IGM), set the neutral
hydrogen fraction in the Ly$\alpha$ forest absorbers, and
help to determine the ion ratios in metal-line absorbers
in QSO spectra.  Ionizing radiation may control the rate of
evolution of the Ly$\alpha$ absorption lines at $z < 2$
(Theuns, Leonard, \& Efstathiou 1998; Dav\'e \ea 1999),
and it may affect the formation rate of dwarf galaxies
(Efstathiou 1992; Quinn, Katz, \& Efstathiou 1996).
The hydrogen photoionization rate, $\Gamma_{\rm HI}(z)$, is
an important component of N-body hydrodynamic modeling
of the IGM.  Because of the large photoionization corrections 
to the observed H~I absorption, the inferred baryon density 
of the IGM and metal abundance ratios also depend on
the intensity and spectrum of this radiation.
Within the Milky Way halo, the ionizing background
can affect the ionization state of high-velocity clouds
located far from sources of stellar 
radiation (Bland-Hawthorn \& Maloney 1999).
 
The ionizing background intensity at
the hydrogen ionization edge ($h \nu_0 = 13.6$~eV)
is denoted $I_0$, in \uvunits, hereafter denoted ``UV units'' or
understood in context. In an optically thin environment, the background
spectrum reflects that of the sources, QSOs and Seyfert galaxies,
which appear to have steep EUV spectra of the form
$F_{\nu} \propto (\nu/\nu_0)^{-\alpha_s}$,
with $\alpha_s = 1.77 \pm 0.15$ from 350--1050 \AA\ (Zheng \ea
1997), or starburst galaxies with $\alpha_s \approx 1.9-2.2$ 
(Sutherland \& Shull 1999).
At high redshift, the IGM is optically thick, owing to the
numerous Ly$\alpha$ absorbers that ionizing photons must traverse.
Thus, the background spectrum is strongly modified by absorption and
re-emission (Haardt \& Madau 1996; Fardal, Giroux, \& Shull 1998, 
henceforth FGS).
 
Estimates of $I_0$ at high redshift are usually obtained
from the ``proximity effect'' (Bajtlik, Duncan, \& Ostriker 1988),
the observed paucity of Ly$\alpha$ absorbers near the QSO
emission redshift.  Recent measurements give values of
$I_0 \approx 10^{-21}$ UV units at $z \approx 3$:
$I_0 = 1.0^{+0.5}_{-0.3} \times 10^{-21}$ (Cooke, Espey, \&
Carswell 1997), $I_0 = (0.5 \pm 0.1) \times 10^{-21}$
(Giallongo \ea 1996), and $I_0 = 0.75 \times 10^{-21}$
(Scott \ea 1999).  At low redshift, the lower comoving
density of QSOs and their diminished characteristic luminosities
suggest that the metagalactic background is reduced
by about a factor of $10^2$ to $I_0 \approx 10^{-23}$.
Using an optical QSO luminosity function (cf. Boyle 1993) and an
empirical model of IGM opacity (Miralda-Escud\'e \& Ostriker 1990),
Madau (1992) estimated that $I_0 =  6 \times 10^{-24}$ 
at $z = 0$. However, this is an uncertain estimate, which now
appears low compared with several local determinations. 
Theoretical extrapolations of $I_0$ to low $z$
are uncertain because they depend sensitively on
the assumed AGN luminosity function and on the IGM opacity
model (Giallongo, Fontana, \& Madau 1997; FGS).
As we will show, the low-$z$ IGM opacity appears to be
dominated by Ly$\alpha$ absorbers in the range 
$14 < \log N\subH < 18$, for which we are just starting to obtain 
statistically reliable information from the {\it Hubble Space
Telescope} (HST).  A future key project on Ly$\alpha$ absorbers
with HST and  a Ly$\beta$ survey with the {\it Far Ultraviolet
Spectroscopic Explorer} (FUSE) should be even more enlightening.
 
Observational estimates of or upper limits on $I_0$ at low redshift have 
been made by a variety of techniques, as described in Table 1.  These 
methods include studies of the proximity effect at 
$\langle z \rangle \approx 0.5$ (Kulkarni \& Fall 1993), edges of H~I
(21 cm) emission in disk galaxies (Maloney 1993; Dove \& Shull 1994a),
and limits on H$\alpha$ emission from high-latitude Galactic clouds
(Vogel \ea 1995; Tufte, Reynolds, \& Haffner 1998) and
extragalactic H~I clouds (Stocke \ea 1991;
Donahue, Aldering, \& Stocke 1995).
Since all these techniques are based on the integrated flux of
LyC radiation, it is convenient to define 
$\Phi_{\rm ion}$ (photons cm$^{-2}$ s$^{-1}$), the normally
incident photon flux through one side of a plane.
For an isotropic, power-law intensity, $I_{\nu} = I_0 (\nu/\nu_0)^{-\alpha_s}$,
we can relate the integral quantity $\Phi_{\rm ion}$ to the
specific intensity $I_0$:
\begin{equation}
   \Phi_{\rm ion} = 2 \pi \int_{0}^{1} \mu \, d \mu \int_{\nu_0}^{\infty}
           \frac {I_{\nu}} {h \nu} \; d \nu
          = \left( \frac {\pi I_0}{h \alpha_s} \right)
            = (2630~{\rm cm}^{-2}~{\rm s}^{-1}) I_{-23}
               \left( \frac {1.8}{\alpha_s} \right)  \;,
\end{equation}
where $\mu = \cos \theta$ is the angle relative to the cloud normal
and $I_{-23}$ is the value of $I_0$ expressed in units of $10^{-23}$
UV units.  Most of the upper limits on $I_0$ translate into values of
$\Phi_{\rm ion}$ in the range $10^4 - 10^5$ photons cm$^{-2}$ s$^{-1}$.
For an assumed EUV spectral index $\alpha_s \approx 1.8$, 
and the approximate form, 
$\sigma_\nu \approx \sigma_0 {(\nu / \nu_0)}^{-3}$, for
the H~I photoionization cross section, the hydrogen
photoionization rate due to this metagalactic intensity is,
\begin{equation}
   \Gamma_{\rm HI} \approx \frac {4 \pi I_0 \sigma_0} {h (3 + \alpha_s) }
        = (2.49 \times 10^{-14}~{\rm s}^{-1}) I_{-23}
       \left( \frac {4.8}{3 + \alpha_s}  \right) \; ,
\end{equation}
where $\sigma_0 = 6.3 \times 10^{-18}$ cm$^2$ is the hydrogen
photoionization cross section at $h \nu_0$.
 
These H$\alpha$ measurements and limits are improving
with better Fabry-Perot techniques (Bland-Hawthorn \ea
1994; Tufte \ea 1998).  In addition, we now have more
reliable HST measurements of the opacity from the low-redshift
Ly$\alpha$ clouds  (Weymann et al. 1998; Shull 1997; Penton, Stocke, 
\& Shull 1999).  Therefore, better computations of the metagalactic 
radiation field are timely.
In this paper, we compute the contribution of Seyfert galaxies,
QSOs, and starburst galaxies to the low-redshift ionizing background 
using three ingredients:  (1) a Seyfert/QSO luminosity function;
(2) AGN fluxes at $\lambda < 912$ \AA\ from extrapolated IUE spectra;
and (3) an improved IGM opacity model, based on recent
HST surveys of Ly$\alpha$ clouds at low redshift.  
In \S~2 we describe these ingredients.
In \S~3 we give the results for $I_0$ and $\Phi_{\rm ion}$ at
$z \approx 0$, together with
error estimates.  In \S~4 we summarize our results and
discuss future work that could improve estimates of $I_0$.

\section*{2. METHODOLOGY } 

The solution to the cosmological radiative transfer equation
(Peebles 1993) for sources with proper specific volume emissivity
$\epsilon(\nu,z)$ (in ergs cm$^{-3}$ s$^{-1}$ Hz$^{-1}$)
yields the familiar expression (Bechtold \ea 1987) for the mean
specific intensity at observed frequency $\nu_{obs}$ as seen by
an observer at redshift $z_{obs}$:
\begin{equation}
   I_{\nu}(\nu_{obs},z_{obs}) = \frac {1}{4 \pi} \int_{z_{obs}}^{\infty}
       \frac {d\ell}{dz} \frac {(1+z_{obs})^3}{(1+z)^3} \;
       \epsilon(\nu,z) \; \exp(-\tau_{\rm eff}) \; dz \; .
\end{equation}
Here, $\nu = \nu_{obs}(1+z)/(1+z_{obs})$ is the frequency of the
emitted photon (redshift $z$) observed at frequency $\nu_{obs}$
(redshift $z_{obs}$), $d\ell/dz = (c/H_0)(1+z)^{-2}(1+\Omega_0z)^{-1/2}$
is the line element for a Friedmann cosmology,
and $\tau_{\rm eff}$ is the effective photoelectric optical depth
due to an ensemble of Ly$\alpha$ absorption systems. For
Poisson-distributed clouds (Paresce, McKee, \& Bowyer 1980),
\begin{equation}
   \tau_{\rm eff}(\nu_{obs},z_{obs},z) 
        = \int_{z_{obs}}^{z} dz' \int_{0}^{\infty}
      \frac {\partial^2 {\cal N}}{\partial N\subH \partial z'}
      \left[ 1 - \exp(-\tau) \right] \; dN\subH \; ,
\end{equation}
where $\partial^2 {\cal N}/\partial N\subH \partial z'$ is the bivariate
distribution of Ly$\alpha$ absorbers in column density and redshift,
and $\tau = N\subH\, \sigma(\nu)$ is the photoelectric (LyC)
optical depth at frequency $\nu$ due to H, He~I, and He~II
through an individual absorber with column
density $N\subH$.  For purposes of assessing the local attenuation
length, it is useful (Fardal \& Shull 1993) to use
the differential form of eq. (4), marking the rate of change of
optical depth with redshift,
\begin{equation}
   \frac {d\tau_{\rm eff}}{dz} = \int_{0}^{\infty}
   \frac {\partial^2 {\cal N}}{\partial N\subH \partial z}
      \left[ 1 - \exp(-\tau) \right] \; dN\subH \;.
\end{equation}
The attenuation length, in redshift units, is given by the
reciprocal of $d\tau_{\rm eff}/dz$.
At low $z$, since $d\tau_{\rm eff}/dz \lsim 1$ at the hydrogen
threshold, its frequency dependence is significant,
and the attenuation length can extend to $z \approx 2$.
In the past few years, more sophisticated solutions to the cosmological
transfer have been developed (Haardt \& Madau 1996; FGS)
taking into account cloud emission and self-shielding.
Figure 1 illustrates our group's recent calculation of the ionizing
background spectrum, computed in full cosmological radiative transfer
with a new IGM opacity model based on high-resolution {\it Keck}
spectra of the Ly$\alpha$ forest and local continua.  These models
include  cloud self-shielding and emission.  We have 
connected this high-redshift opacity model with our new
model from HST studies (discussed below) for the low-redshift opacity
at a transition redshift $z = 1.9$.  By redshift $z = 0$, 
the intensity has declined to $I_0 \approx 1.3 \times 10^{-23}$, 
corresponding to $\Phi_{\rm ion} \approx 3000$ photons cm$^{-2}$ s$^{-1}$ 
for sources with $\alpha_s \approx 2$.

In the work that follows, we compute $I_{\nu}$ using both our
detailed cosmological radiative transfer code and an
approximate solution to equation (3).  In this approximation,
we neglect the effects of emission from attenuating absorbers and 
approximate the opacity with a simple power-law fit that neglects
the effects of He absorption.  We will discuss the accuracy
of this approximation in more detail in \S~3.  
Because the opacity of the IGM is much smaller at low
redshift, this more rapid calculation is adequate 
for estimating the present-day level of radiation just above $\nu_0$.
The primary ingredients for the computation of $I_0$ at low redshift
are the source emissivity, $\epsilon(\nu,z)$, and the opacity model for
$\tau_{\rm eff}(\nu_{obs},z_{obs},z)$. In the following sub-sections, we
describe how we determined these quantities.

\subsection*{2.1. AGN Luminosity Function and Spectra}

The distribution of AGN luminosities is typically described by a
rest-frame, B-band luminosity function.  In order to estimate the total
emissivity of AGN at the Lyman limit, we must know both the 
luminosity function and the average spectrum of the AGN.  In addition,
we must know the assumptions about the spectrum that were made to
construct the luminosity function.

To estimate the intrinsic AGN quasar spectrum, we begin with the
Seyfert optical sample of Cheng \ea (1985), based on Seyfert 1 and 1.5
galaxies covered by the first nine Markarian lists.  Their sample was
corrected for incompleteness, and the contribution from the host galaxy
was subtracted out.  The separation of nuclear and host galaxy
luminosity becomes increasingly challenging at the faint end of the
luminosity function.  Ideally, careful, small-aperture photometry
would be used for these estimates.  Cheng \ea (1985) relied instead on
two independent methods to separate the contribution from the nucleus.
In the first method, they assumed a template host galaxy and corrected
this for orientation and internal extinction.  In the second method,
they assumed that all nuclei had the same intrinsic colors and
determined the nuclear contribution via the color-given method of
Sandage (1973).  They found these two methods to give consistent
results.  In addition, they compared the color-given method with
nuclear magnitudes derived by careful surface photometry for 11
Seyfert 1 galaxies (Yee 1983).  They assigned a total uncertainty of
$0.5$ mag in the nuclear $M_B$ to the sample.  On re-examination of
the sample, we found that the errors most likely decrease with the
luminosity of the galaxy.  We assume that the errors on the specific
luminosity, $L_B$ (ergs s$^{-1}$ Hz$^{-1}$), decrease linearly from 
0.24 dex at $\log L_B=28$ to 0.16 dex at $\log L_B=30$.

From this sample of Seyferts, we chose 27 objects observed repeatedly
by the {\it International Ultraviolet Explorer} (IUE) satellite.
Together with many other AGN, these Seyfert galaxies are part of the 
Colorado IUE-AGN database (Penton, Shull, \& Edelson 1996), which gives 
both mean and median spectra. Since these AGN are subject to flux
variability, the distribution in flux is a skewed distribution with
a tail that includes short flares studied by various IUE campaigns. 
To provide a conservative estimate of the ionizing fluxes, we have 
therefore used median spectra to derive correlations;  however the
differences in the correlations are only a few percent. The
line-free regions of the median IUE spectra were fitted to power-law
continua and extrapolated to 912 \AA\ (rest-frame), from which we
derive the specific luminosity, $L_{912}$ (ergs s$^{-1}$ Hz$^{-1}$).
We also convert from $M_B$ to $L_B$, at $\nu = 6.81 \times 10^{14}$ Hz 
(4400 \AA) by the formula derived from Weedman (1986, eqs. 3.15 and 3.16), 
$\log L_B=0.4(51.79-M_B)$.
Figure 2 shows the correlation between $L_{912}$ and $L_B$.  The error
bars are only shown for $L_B$, since the errors in $L_{912}$ are much
smaller.  We find that $L_{912} = (2.60 \pm 0.22) \times 10^{28}
(L_B/10^{29})^{(1.114 \pm 0.081)}$ erg s$^{-1}$ Hz$^{-1}$. 
This translates to a UV-optical spectral slope that depends on $L_B$ as
\begin{equation}
\label{slope-luminosity}
\alpha_{\rm UV} = (0.86 \pm 0.05) - (0.16 \pm 0.12) \log(L_B/10^{29}) \; .
\end{equation}
The evidence for $L_B$ dependence is marginal, however, and our basic
model will simply assume a constant slope $\alpha_{\rm UV} = 0.86$ between the
B band and the Lyman limit.  This agrees quite well with the average
QSO spectrum derived by Zheng \ea (1997).

The second ingredient in the computation of the AGN emissivity is the
luminosity function.  The function that matches observations over the
broadest redshift and luminosity range is the comoving
analytic form given by Pei (1995, eqs. 6--8),
\begin{equation}
   \Phi(L,z) = \frac { \Phi_{\ast} / L_z}
       {(L/L_z)^{\beta_l} + (L/L_z)^{\beta_h} } \; ,
\end{equation}
where the characteristic ``break luminosity'' is given by
\begin{eqnarray}
   L_z &=& L_* (1+z)^{-(1-\alpha_{\rm UV})} \exp \left[ - (z-z_*)^2/2\sigma_*^2
             \right] \nonumber \\
        &=& L_0 (1+z)^{-(1 - \alpha_{\rm UV})} \exp \left[
            -z(z-2z_*) / 2 \sigma_*^2 \right] \; .
\end{eqnarray}
Here, $\alpha_{\rm UV}$ is the UV-optical spectral index and $L_0 = L_*
\exp(-z_*^2/2 \sigma_*^2)$ is the ``break luminosity'', $L_z$, at the
present epoch.  Note that we define our spectral index with
the opposite sign to that assumed by Pei, i.e., we adopt 
$L_\nu \propto \nu^{-\alpha_{\rm UV}}$ where most data suggest that 
$0.5 < \alpha_{\rm UV} < 1$.  We present results for luminosity functions based 
on the two sets of assumptions about the cosmology and optical spectral
index $\alpha_{\rm UV}$ derived by Pei.  The ``open model'' has $h = 0.5$,
$\Omega_0=0.2$, and $\alpha_{\rm UV} = 1.0$ and yields $\beta_l = 1.83$,
$\beta_h=3.70$, $z_* = 2.77$, $\log (L_*/L_\odot)=13.42$,
$\sigma_{\ast} = 0.91$, and $\log (\Phi_{\ast} /{\rm Mpc}^{-3})=-6.63$.  
The ``closed universe'' model has $h = 0.5$, $\Omega_0=1$, and 
$\alpha_{\rm UV} = 0.5$ 
and yields $\beta_l = 1.64$,
$\beta_h=3.52$, $z_* = 2.75$, $\log (L_*/L_\odot) = 13.03$,
$\sigma_{\ast} = 0.93$, and $\log (\Phi_{\ast} /{\rm Mpc}^{-3})=-6.05$.  
Figure 3 shows these two models at $z=0$.  After
correcting for the different spectral indices, there is no physical
reason why these two models should differ in the ionizing intensity
they imply or in their value as $z \rightarrow 0$. However, the
intensity estimates from these models differ by up to $40\%$ (see
FGS).  This points to substantial uncertainties in the Pei fit that
are not reflected in the small formal errors.

We integrate the ionizing luminosity density over the luminosity
function from $L_{\rm min}$ to $L_{\rm max}$.  In our standard
model, we assume that these limits are $0.01 L_z$ and $10 L_z$, 
respectively, and we explore the sensitivity of the results to
$L_{\rm min}$.  If we define $x = L/L_z$, we can write the comoving
specific volume emissivity as
\begin{equation}
   \epsilon(\nu,z)= \Phi_{\ast}
      \left( \frac{\nu_{0}}{\nu_{B}} \right)^{-\alpha_{\rm UV}}
      \left( \frac{\nu}{\nu_{0}} \right)^{-\alpha_{s}}  L_z
    \int_{x_{\rm min}}^{x_{\rm max}} \frac
   {x}{x^{\beta_l}+x^{\beta_h}} \; dx \; ,
\end{equation}
where we assume an EUV power-law spectral index $\alpha_s$ for the AGN
and integrate from $x_{\rm min} = L_{\rm min}/L_z$ to $x_{\rm max} =
L_{\rm max}/L_z$. With $x_{\rm min} = 0.01$ and $x_{\rm max} = 10$,
our results are insensitive to $x_{\rm max}$ and moderately
insensitive to $x_{\rm min}$.  An increase (decrease) of $x_{\rm min}$
by a factor of 3 leads to a decrease (increase) of the calculated
emissivity by $15\%$.  A similar change of $x_{\rm max}$ changes the
emissivity by only 2\%.  From several trials, we find an
empirical scaling relation,
$I_0^{\rm AGN} \propto (L_{\rm min}/0.01 L_*)^{-0.17} (\alpha_s/1.8)^{-0.97}$. 
To better assess the uncertainty from the
extrapolation to bright and faint sources, we use Figure 2 of Pei
(1995), which shows the range of B luminosities that contribute to the
luminosity function fit.  We define the ``completeness'' as the
integral of the emissivity over this range, compared with the integral
for the standard range $0.01 < x < 10$.  This completeness rises
from 20\% at $z=0$ to 80\% at $z \sim 2$, but falls to 20\% at $z
\sim 4$.  The low-$z$ results can be made more robust by
considering additional surveys of Seyferts and QSOs.
 
Figure 3 shows the B-magnitude luminosity function from the Cheng \ea 
(1985) sample, as well as the results from the survey of K\"{o}hler
\ea (1997) for Seyferts and QSOs with $z<3$.  These results are also 
compatible with the results of the local optical luminosity function 
based on an X-ray selected sample of AGN (Della Ceca \ea 1996).
There are some discrepancies, however,
between these results and the Pei luminosity function.  There are more
QSOs at the bright end of the luminosity function 
at $z < 1$ than in the Pei
form, an effect that has been noted by several authors 
(Goldschmidt \ea 1992; Goldschmidt \& Miller 1998; La Franca \&
Cristiani 1997).  This appears to have its origin in systematic errors
in the Schmidt \& Green (1983) QSO survey.  In addition, there appears
to be a slight deficit at the knee of the Pei function in both the
K\"{o}hler \ea and Cheng \ea samples.  In fact, the K\"{o}hler results
are fitted adequately with a single power law, as shown in Fig. 3.
 
It turns out that these two effects cancel to within $5\%$ when we
compute the total B-band emissivity over the range $0.01 < L_B/L_z(0)
< 10$.  However, the fact that the luminosity function appears to be
changing shape weakens the assumptions leading to the Pei function,
and the convergence at the faint end becomes even slower.

Hence, our best estimate of the emissivity is unchanged from the Pei
(1995) model, but we suspect that there are still substantial
uncertainties in the AGN luminosity function.  In Fig. 4 we plot 
the emissivity from the Pei ``open'' model.
To find a reasonable range of emissivity models, we
multiply (or divide) this by the square root of the ``completeness'' of
the Pei model as derived above, and we also change the spectral slope
$\alpha_{\rm UV}$ and its dependence on $L_B$ by one standard deviation to
minimize (or maximize) the ionizing emissivity.  We consider the band
thus defined to be a conservative 1$\sigma$ range for the
emissivity.  Note that the Pei ``closed'' model, adjusted for the
different spectral index and cosmology, lies well within this range.

The average UV QSO spectrum derived by Zheng \ea (1997) implies
an EUV spectral index $\alpha_s = 1.77 \pm 0.15$ for radio-quiet 
QSOs and $\alpha_s = 2.16 \pm 0.15$ for radio-loud quasars. 
Since radio-quiet QSOs are much more numerous, we select
$\alpha_s = 1.8 \pm 0.15$ as a representative measure of the EUV 
spectral index.
 
\subsection*{2.2.  Stellar Contributions to the Ionizing Emissivity}

The contribution of stars within galaxies to the ionizing background 
remains almost completely unknown at all epochs.  It has been realized
(cf. Madau \& Shull 1996) that the amount of ionizing
photons associated with the production via supernovae
of the observed amount of metals in the universe might
easily exceed the ionizing photons produced by AGNs.
The problem has been to estimate the average fraction,
$\langle f_{esc} \rangle$, of ionizing photons 
produced by O and B stars that escape the galaxies into the IGM.
Thus far, observational limits on $\langle f_{esc} \rangle$
exist from only a few nearby galaxies (Leitherer \ea 1995; 
Hurwitz \ea 1997).  On theoretical grounds, Dove \& Shull (1994b) 
concluded that an escape fraction of order 10\% might be possible,
while more recent models of the escape of ionizing photons through
supershell chimneys (Dove, Shull, \& Ferrara 1999) suggest 
fractions of 3--6\%, which are compatible with the observational 
limits.  Recently, Bland-Hawthorn \& Maloney (1999) 
used measurements of H$\alpha$ from the Magellanic Stream
to infer an escape fraction $\sim$ 6\% from the Milky Way.

Deharveng \ea (1997) argued that even the present
indirect upper limits on $I_0$ must limit the
present-day escape fraction to below 1\%. 
They depart radically, however, from the
work of Dove \& Shull (1994b) and Dove et al. (1999)
in their treatment of the nature of the escape fraction.
Deharveng \ea assume that stellar ionizing photons are prevented 
from escaping their host galaxies by the opacity of neutral hydrogen, 
effectively in the limit that the hydrogen forms an unbroken sheet.
Consequently, the opacity varies approximately as ${(\nu/\nu_0)}^{-3}$,
so that the escape of higher-energy photons is dramatically enhanced.  
For example, the optical depth can decrease from
$\tau \approx 50$ at the Lyman edge to $\tau < 1$
at 4 rydbergs.  For photons emitted at high redshift that 
survive to contribute to $I_0$ at $z = 0$, the sharp increase in the
escape fraction outweighs the effects of redshifting.
In the Deharveng picture, the contribution of high-$z$ galaxies to 
the present-day mean intensity at $\nu=\nu_0$ is strongly enhanced.  
Thus, if $\langle f_{esc} \rangle \approx 0.001$,
starbursts may match the contribution of quasars to the
present-day ionizing mean intensity.  

However, we believe the Deharveng model for photon escape is
physically unrealistic.   Our alternative view is that
the internal galactic opacity to all
stellar photons is large, and photons may
only escape from isolated regions of high star
formation whose H~II regions or attendant
supershells are able to break through this high
opacity layer (Dove et al. 1999).
Within this view, a constant escape fraction with
frequency is more appropriate.

Our estimate of the stellar ionizing photons is made in the 
following way.  Gallego \ea (1995) performed an H$\alpha$
survey of galaxies and fitted their derived luminosity
function to a Schechter function.  By integrating this,
they were able to estimate a total H$\alpha$
luminosity per unit volume at low redshift, 
$L_{H\alpha} = 10^{39.1\pm 0.2}$ erg~s$^{-1}$ Mpc$^{-3}$.
In the usual fashion, we relate this to the number of ionizing photons
by dividing the number of H$\alpha$ photons by the
fraction of the total H recombinations that produce
an H$\alpha$ photon.  We multiply their estimate of the
ionizing photon production by $\langle f_{esc} \rangle$ to give the 
total ionizing photons in the IGM.   The representative models assume
$\langle f_{esc} \rangle = 0.05$ and are shown in Figure 5.
Preliminary results are now available from the KPNO International 
Spectroscopic Survey (KISS) (Gronwall et al. 1998), which probes to 
fainter magnitudes than the Gallego \ea survey.  
Gronwall (1998) quotes a value 
$L_{H\alpha} = 10^{39.03}$ erg~s$^{-1}$~Mpc$^{-3}$, but notes that
these results are preliminary and represent a lower limit to
the true H$\alpha$ density, since even their deeper survey is 
incomplete for galaxies with H$\alpha$ emission-line equivalent 
widths less than 25 \AA.  

The spectrum of ionizing photons from clusters of hot stars differs 
from that of AGN in that stars emit relatively few photons more energetic 
than 4 Ryd. Sutherland \& Shull (1999) have shown that, between
1 and 4 Ryd, the spectrum of a starburst may be
approximated as a power law with spectral index $\alpha_s = 1.9 - 2.2$.

We have tied the redshift evolution in ionizing photon
production rate to the star formation evolution
observations of Connolly \ea (1997) based upon
the Hubble Deep Field (HDF).  The effects of
dust extinction remain a major uncertainty
in the determination of the star formation rate
at high redshift (cf. Pettini \ea 1998). Many high-$z$
star-forming galaxies appear to be dust-obscured, based
on recent sub-millimeter studies of the HDF (Hughes et al.
1998; Barger et al. 1998).   In addition, a survey of Lyman-break galaxies 
at $z \approx 3-4$ (Steidel \ea 1999), covering a much larger 
angular extent than the HDF, finds no significant difference
in the star formation rate at $z=3$ and $z = 4$.   
With corrections for dust extinction,
the star formation could remain constant from $z = 1.5$ out to
$z > 4$.  As a result, we have also considered a case 
in which the star formation remains constant after
reaching its peak at $z \approx 2$.  However, despite the
many uncertainties of high-$z$ star formation rates,
the effects on the present-day level of ionization
are minimal, of order a few percent.

\subsection*{2.3.  Absorption Model for the IGM } 

At $z < 2$, the redshift densities of the \Lya\ clouds and Lyman limit
systems decline steeply with cosmic time.  Morris \ea (1991) and
Bahcall \ea (1991) used HST observations to show that this decline
could not be extrapolated to the present, as far too many
\Lya\ absorbers were observed toward 3C~273.  The large
data set gathered by the Hubble Key Project with the
Faint Object Spectrograph (FOS) 
shows a sharp break in the evolving redshift density
at $z \sim 1.5-2$ (Weymann \ea 1998). Ikeuchi \& Turner (1991)
showed that the cessation in this steep decline was a natural 
consequence of the falloff in the ionizing emissivity from
$z=2$ down to  0.  This conclusion has been
borne out by detailed cosmological simulations (Dav\'e \ea 1999),
which indicate that the effect is insensitive to the specific
cosmological model. 

In our calculations, we will base our absorption model on observations.
Our analysis follows the traditional ``line-counting'' method, where
spectral lines are identified by Voigt profile-fitting and the
opacity is calculated by assuming a Poisson distribution of these
lines.  At high redshift, FGS used this method to
estimate the opacity based on high-resolution observations of QSO
Ly$\alpha$ absorption lines from {\it Keck} and other large-aperture
telescopes in the redshift range $2 \leq z \leq 4$.  It is not appropriate
to extrapolate this model to lower redshifts, owing to 
the rapid evolution rate of the Ly$\alpha$ forest.

Our method of determining $d\tau_{\rm eff}/dz$ considers Ly$\alpha$ lines 
in three ranges of column density:  from $12.5 < \log N\subH < 14.0$
(HST/GHRS survey), from $14.0 < \log N\subH < 16$ (HST/FOS survey),
and for $\log N\subH > 17$ (HST/FOS Lyman-limit survey).  
The HST/FOS survey forms the core of our standard opacity model.
We combine these results with HST/GHRS measurements of weak lines and with 
the HST/FOS survey of Lyman-limit systems, extrapolated downward from 
the Lyman limit by two different methods.   
At redshifts $z < 1.5$, the most extensive study of strong Ly$\alpha$
absorbers ($10^{14 - 16}$ cm$^{-2}$) is the QSO Absorption Line
Key Project with HST/FOS (Januzzi \ea 1998; Weymann \ea 1998).
Weaker Ly$\alpha$ lines, which contribute a
small amount to the opacity and serve as a constraint on the column density
distribution, were studied by Shull (1997), Shull, Penton, \& Stocke (1999),
and Penton \ea (1999) using HST/GHRS spectra.
The distribution of Lyman limit systems with $N\subH > 10^{17}$ cm$^{-2}$
are discussed by Stengler-Larrea \ea (1995) and Storrie-Lombardi \ea (1994).
Each of these surveys suffers from incompleteness or saturation effects in
various regimes. Therefore, extrapolations outside the range of 
$N\subH$ and comparisons in regimes of overlap are helpful.
Extensive future UV surveys with HST and FUSE will also
reduce some of the uncertainties (see discussion in \S~4).

HST/GHRS studies of Ly$\alpha$ absorbers in the range $12.5 \leq
\log N\subH \leq 14.0$ (Penton \ea (1999) find a column density
distribution, $d{\cal N}/dN\subH \propto N\subH^{-1.74 \pm 0.26}$.
The cumulative opacity of these weak lines (up to $10^{14}$ cm$^{-2}$)
is relatively small:  
$d\tau_{\rm eff}/dz = 0.025 \pm 0.005$ for the low-redshift range
($0.003 < z < 0.07$). However, a small number of higher column density
systems produce a steady rise in the cumulative opacity for
$\log N\subH > 14$.  Extending the HST/GHRS distribution up to 
$\log N\subH = 15$ gives
$d\tau_{\rm eff}/dz \approx 0.09 \pm 0.02$. Above this column density,
saturation effects and small-number statistics make the number
counts more imprecise. In the range $15 < \log N\subH < 16$, Penton
\ea (1999) estimate an additional contribution,
$d\tau_{\rm eff}/dz \approx 0.1 - 0.3$.

The HST/FOS Key Project spectra have insufficient resolution to determine 
line widths or to resolve velocity components.  As a result, the conversion
from equivalent width, $W_{\lambda}$, to column density, $N\subH$, is
difficult for saturated lines.  
In the absence of other lines (e.g., Ly$\beta$), one can only estimate the
conversion from $W_{\lambda}$ to $N\subH$ by assuming a curve of
growth and doppler parameter.
This difficulty was noted by Hurwitz \ea (1998)
in their comparison of unexpectedly strong ORFEUS Ly$\beta$ absorption 
compared to predictions from HST Ly$\alpha$ lines toward 3C~273. 
For unsaturated lines, $W_{\lambda} = (54.4~{\rm m\AA}) N_{13}$, 
where $N\subH = (10^{13}~{\rm cm}^{-2}) N_{13}$ and where the 
line-center optical depth is $\tau_0 = (0.303) N_{13} b_{25}^{-1}$ 
for a doppler parameter $b = (25~{\rm km~s}^{-1}) b_{25}$.  The 
HST/FOS Key Project lines with $W_{\lambda} \geq 240$ m\AA\
are highly saturated in the range ($14 \leq \log N\subH \leq 17$) that
dominates the Ly$\alpha$ forest's contribution to the continuum opacity. 

As a first attempt to incorporate the Key Project information, we
focus on the statistical frequency of Ly$\alpha$ absorbers, 
$d {\cal N}/dz = 30.7 \pm 4.2$, for lines with $W_{\lambda} > 240$ m\AA.  
For $b = 25~{\rm km~s}^{-1}$, this corresponds roughly to $\log N\subH > 14$.  
We compute opacities based upon sample 5 of Weymann \ea (1998),
which includes 465 absorption lines ($W_\lambda > 240$ m\AA) that
could not be matched with corresponding metal lines.  This sample was
intended to remove high column density lines that may evolve more rapidly
with redshift, consistent with the results of Stengler-Larrea \ea
(1995) for Lyman limit systems.  This segregation has little
effect on the opacity.  The Key Project became incomplete at
column densities well below the Lyman limit.  Therefore, to determine
an IGM opacity, we needed to extrapolate to $\log N\subH = 17$
using assumptions about the distribution. 
 
The bivariate distribution of Ly$\alpha$
absorbers per unit redshift and unit column density can be expressed as
$\partial^2 {\cal N}/\partial z\, \partial N\subH
= A (N\subH/10^{17} \, {\rm cm}^{-2})^{-\beta} (1+z)^{\gamma}$.
Improvements on this form have been suggested, notably
the addition of one or more breaks in the power-law distribution 
(Petitjean \ea 1993; FGS).  Here, we parameterize our uncertainty
by assuming just one power-law index, but we vary the upper limit
on column density to which we integrate.
Figure 6 shows a set of curves, corresponding to $\beta = 1.5 \pm 0.2$,
of the differential effective opacity, $d\tau_{\rm eff}/dz$, 
evaluated at $z = 0$ and at the hydrogen threshold.  
Assuming $\beta = 1.5$ and no break in the distribution, 
we find $d\tau_{\rm eff}/dz \approx 0.2$ for the FOS range
$10^{14-16}$ cm$^{-2}$ and $d\tau_{\rm eff}/dz \approx 0.5$ 
for the expanded range $10^{14-17}$ cm$^{-2}$. 
Because $\gamma = 0.15 \pm 0.23$ for this FOS sample, the opacity 
does not change greatly with redshift.  We choose a standard value 
$d\tau_{\rm eff}/dz \approx 0.5$, corresponding to $\beta=1.5$ and 
$N_{\rm max} = 10^{17}~~{\rm cm}^{-2}$.  

The Lyman-limit data (Stengler-Larrea \ea 1995) can be fitted to the form
\begin{equation}
  \frac {d \tau_{\rm eff}}{dz} = 0.263 (1+z)^{1.50} \; ,
\end{equation}
assuming that $\partial^2 {\cal N}/\partial z\, \partial
N\subH \propto N\subH^{-1.5} (1+z)^{1.50}$.  In absorption model 1, we
took the lower limit on $N\subH$ for partial LL systems to be $N_{l} =
10^{17}$ cm$^{-2}$.  If this limit is extended down to $10^{16}$
cm$^{-2}$ or $10^{15.5}$ cm$^{-2}$, the coefficient 0.263 in eq. (10)
increases to 0.382 and 0.411, respectively; the latter choice becomes
our Model 2.

We summarize our three opacity models in Table~2.  We use equation~(5)
to calculate the opacity, but include an approximation
of the frequency dependence of the opacity over the range 1--3 Ryd,
which dominates the H~I photoionization rate.  This approximation
takes the form
\begin{equation}
   \frac {d \tau_{\rm eff}}{dz} = c_i \left(
     \frac {\nu}{\nu_0} \right) ^{s_i}  (1+z)^{\gamma_i} \; .
\end{equation}
In Figure 7, we compare $d\tau_{\rm eff}/dz$ at $\nu = \nu_0$ for the
three models described above, as well as low-redshift extrapolations
of the opacities of FGS and Haardt \& Madau (1996).  It
can be seen that the poorly determined column-density distribution of the
\Lya\ forest leaves a large uncertainty in the total opacity, even
though the evolution of the number density is tightly constrained by
the HST observations.  The partial LL systems ($16 \leq \log N\subH \leq 
17.5$) probably dominate the IGM opacity
at low redshift, but they are so rare that statistical fluctuations
from sightline to sightline are quite large.  It will require many 
high signal-to-noise spectra along
low redshift lines of sight to reduce the uncertainty in
their contribution to the opacity.

\section*{3.  RESULTS: THE IONIZING RADIATION FIELD }

\subsection*{3.1. The Contribution from AGN}

Our best-estimate model for the present day intensity $I_0$
makes the following four assumptions:
(1)  The AGN distribution is described by the Pei (1995)
QSO luminosity function with $h = 0.5$ and $\Omega_0 =
0.2$, modified by the assumption that the optical-UV
spectral index $\alpha_{\rm UV} = 0.86$, but with no
correction for intervening dust.
(2)  The lower (upper) cutoffs to the luminosity
function are $L_{\rm min(max)} = 0.01 (10) L_*$.
(3)  The opacity below $z \approx 1.9$ is our
``standard model,'' itself based upon HST observations
of Ly$\alpha$ forest and Lyman Limit systems.  Above $z \approx 1.9$, 
it is Model A2 from FGS.
(4)  The ionizing spectrum ($\nu \geq \nu_0$) has spectral index
$\alpha_s = 1.8$.  Using the full radiative transfer
calculation outlined in FGS, we find
\begin{equation}
   I_0 = 1.3^{+0.8}_{-0.5} 
         \times 10^{-23}~{\rm erg~cm}^{-2}~{\rm s}^{-1}~
         {\rm Hz}^{-1}~{\rm sr}^{-1}.     
\end{equation}
As we discuss below, most of the uncertainties in our
estimate are essentially multiplicative.  The
uncertainties quoted above represent the addition
in quadrature of the uncertainties 
in the log of the factors discussed below.
Direct addition in quadrature of the relative uncertainties
gives a similar range of uncertainty.
Strictly speaking, although none of the uncertainties is 
independent, we treat them as if they were in computing the 
total uncertainty.

Figure 4 summarizes the uncertainties in the emissivity due to
AGN that pertain to assumptions (1) and (2).
The corresponding 1 $\sigma$ uncertainty in the calculated specific 
intensity for the $\Omega_0 = 0.2$ case is $\pm 0.19$ dex.
As FGS noted, the intensity should be independent
of the choice of $H_0$ or $\Omega_0$ if the
emissivity is actually determined observationally.
The AGN luminosity function is
propagated from low to high redshift, assuming
pure luminosity evolution, a parameterization
that depends only on $z$, and not explicitly
on appropriate cosmological distances and
luminosity evolution parameters.  The
difference in the calculated mean intensity
for a closed universe ($\Omega_0=1$) is not
an independent source of uncertainty.
However, we estimate an uncertainty of $\pm 0.01$ dex, 
which is negligible.

The uncertainties inherent in our choice of opacity model appear
both in the degree of attenuation of ionizing photons due to the 
absorbers, and in the contribution of
diffuse ionizing radiation from the absorbers
(mainly He~II ionizing photons reprocessed to He II Ly$\alpha$
and two-photon radiation).  Using Models 1 and 2 instead of our 
standard model, we calculate slightly higher levels of $I_0$,
by factors of $0.0068$ and $0.0329$ dex.  A sample
standard deviation of the models is $\pm 0.017$ dex
and is probably an adequate assessment of
the $1 \sigma$ uncertainty in $I_0$ due to the
opacity.  A more complete picture of the
full (at least several $\sigma$) uncertainty 
may be obtained from the following extreme cases.
The diffuse ionizing radiation from absorbers
contributes 20\% of the ionizing intensity at
$z = 0$ (see Fig. 8a), so eliminating this contribution reduces
$I_0$ by $0.1$ dex.   If there is no opacity at all, $I_0$
increases by $0.26$ dex.  If the column density
distribution for the absorbers in our standard
model is assumed have an unbroken power law with
$\beta = 1.3$ rather than 1.5, the reduction in $I_0$ is $0.21$ dex.
The shape of the spectrum of AGN shortward of
$912$ \AA\ is a relatively small source of uncertainty.  
As $\alpha_s$ is varied between 1.5 and 2.1, 
representative of the 2 $\sigma$ uncertainty in $\alpha_s$, the relative
specific intensity varies by $\pm 0.1$ about 
the standard value $\alpha_s = 1.8$.   We therefore
adopt a conservative uncertainty of $\pm 0.1$ dex in $I_0$. 

There remains an additional systematic uncertainty in the contribution 
of AGN to $I_0$.  As discussed by FGS, the Pei luminosity function 
used here produces insufficient ionizing photons to account 
for the level of the ionizing background at $z > 3.5$
implied by the proximity effect.  Further,
there are not enough ionizing photons to 
reionize the universe by $z \approx 5$ (Madau 1998), the epoch
of the highest redshift quasars.  While the latter point
remains a problem for scenarios in
which only AGN ionize the IGM, the former difficulty
is ameliorated by the suggestion
that, especially at higher redshifts, the number
of AGN are being undercounted due to obscuration by dust-laden 
absorbers (Heisler \& Ostriker 1988; Fall \& Pei 1993; Pei 1995).  
Using the dust-corrected luminosity function of Pei (1995),
we find that $I_0$ at $z = 0$ is increased by $0.08$ dex.

As shown in Fig. 1, the level of the mean intensity at $z \approx 2$ 
calculated by FGS is $I_0 \approx 7 \times 10^{-22}$.
For this model, ionizing radiation due to sources at
$z > 2$ produces 20\% of the mean intensity at $z = 0$ (see Fig. 8a).
As Fig. 1 shows, redshifted He~II Ly$\alpha$ diffuse emission is 
still substantial at $z = 2$.  Because He~II $\lambda304$ 
emitted at $z > 2$ is redshifted below the H~I threshold by $z = 0$,
this emission contributes less than 10\% at $z=0$.  
To give a specific example, suppose that
the number of AGN at high redshift was severely underestimated,
so that the metagalactic background due to AGN at $z \approx 2-3$
was quintupled, while retaining the same spectral shape.  Then, 
the value of $I_0$ at $z=0$ would be increased by only $0.15$ dex.
Even for AGN, the strong attenuation of high-energy
photons by He~II absorption greatly limits any
contribution to the present-day ionizing background
from sources at $z > 3$.  A significantly larger population of AGN
at $z = 2-3$ will not augment $I_0$ by more
than 40\% at $z = 0$.  This systematic effect
has not been included in the uncertainty in eq. (12).

\subsection*{3.2. The Possible Contribution from Hot Stars}

An estimate of the ionizing radiation contributed
by hot stars is complicated by its dependence on
factors such as $\langle f_{esc} \rangle$,
for which a good estimate of its magnitude and uncertainty 
does not exist.  As a result, we include these factors
explicitly in our results.  Our model
for the possible contribution of stars to the
present day specific intensity $I_0$ makes
the following assumptions:  (1)  The production
of ionizing photons by stars at the present time
may be calibrated by the density
of H$\alpha$ photons in the extragalactic background.
(2)  The star formation rate is proportional to that from the 
observations of Connolly \ea (1997) and assumes $h = 0.5$
and  $\Omega_0 = 0.2$.  
(3)  The IGM opacity is our ``standard model.'' 
(4)  The average spectrum of the OB associations
that provide the ionizing photons has spectral index
$\alpha_s = 1.9$ (1 -- 4 Ryd) with no photons
above 4 Ryd. Using a radiative transfer calculation that neglects 
the diffuse radiation contributed by intervening absorbers 
(see discussion below), we find
\begin{equation}
   I_0 = 1.1^{+1.4}_{-0.7} \times 10^{-23}
         \left( \frac {\langle f_{esc} \rangle} {0.05} \right) 
         \left( \frac {L_{H\alpha} } {10^{39.1}} \right) 
         {\rm erg~cm}^{-2}~{\rm s}^{-1}~
         {\rm Hz}^{-1}~{\rm sr}^{-1}    \; , 
\end{equation}
where we have scaled the H$\alpha$ luminosity density, 
$L_{H\alpha}$, to the Gallego et al. (1995) standard value, 
$10^{39.1}$ erg s$^{-1}$ Mpc$^{-3}$, and adopted a probable LyC
escape fraction, $\langle f_{\rm esc} \rangle = 0.05$.

The uncertainties for which we have not directly
parameterized our ignorance are mainly mutiplicative, as in the
case for AGN. The uncertainty in the numerical coefficient
quoted above again represents the addition
in quadrature of the uncertainties in the log of the factors
discussed below.  We make the same conservative assumption that the
individual uncertainties are independent.
The uncertainty in the H$\alpha$ luminosity density suggested
by Gallego \ea (1995) is $\pm 0.2$ dex, while
the uncertainty in the preliminary KISS result, is $\pm 0.1$ dex. 
We adopt an uncertainty of $\pm 0.2$ dex for the H$\alpha$
calibration, but it is possible that this
uncertainty will be reduced substantially
when full details of KISS are released. As long as
$\langle f_{esc} \rangle$ is small, so that
recombinations within the galaxies provide a fair accounting 
of the number of ionizing photons produced,
this is the uncertainty we associate with assumption
(1).  The conversion of H$\alpha$ photons to
ionizing photons is nearly temperature independent,
so uncertainties in the temperature of ionized
regions in galaxies are negligible.  The 
H$\alpha$ emissivity is parameterized directly
in units of the luminosity density suggested by
Gallego \ea (1995), but the uncertainty discussed
above is included in the numerical coefficient.

The uncertainties in the stellar emissivity 
(assumption 2) are summarized
in Fig. 5.  The error bars in the points 
from the Connolly \ea data are formulated
in a different way from our evaluation of the AGN
uncertainty.  These authors corrected
for survey incompleteness using a 
Schechter function with three different power laws
to extrapolate to low luminosity.
We show a rough $1 \sigma$ range in the emissivity
based upon this estimate of the uncertainties.
Our emissivity depends upon a fit to the data
points in Fig. 5, and has small
variations with the assumed cosmology.  This introduces
a small uncertainty of 0.01 dex in the calculation of $I_0$.

The uncertainty inherent in our choice of
opacity model, assumption (3) above, appears
primarily in the degree of IGM attenuation of ionizing
photons.  If we focus only on the stellar contribution to the ionizing
background, the number of He~II ionizing photons produced is negligible. 
As a result, the contribution of diffuse ionizing
radiation from the absorbers may be neglected.
(Because H~I recombination radiation is closely confined
to the ionization edge, this diffuse radiation is quickly
redshifted below $\nu_0$.) Using Models 1 and 2 
for the opacity instead of our standard model,
we calculate slightly higher levels of $I_0$,
by factors of 0.0068 and 0.0329 dex.  A sample
standard deviation of the models is $\pm0.017$ dex
and is probably an adequate assessment of
the $1 \sigma$ uncertainty in $I_0$ due to the opacity.  
If there is no opacity at all, $I_0$
increases by $0.26$ dex.  If the column density
distribution for the absorbers in our standard
model has an unbroken power law, with $\beta = 1.3$ rather than
1.5, the reduction in $I_0$ is $0.21$ dex.

Our assumption (4), that the stellar ionizing radiation has 
a spectrum $\nu^{-1.9 \pm 0.2}$ (Sutherland \& Shull 1999), yields 
a relatively small source of uncertainty.  As $\alpha_s$
is varied between $1.7-2.1$, the 
specific intensity increases by factors of $0.045$ to $-0.043$ dex.
We therefore assign an uncertainty of $\pm0.045$ dex.

As shown in Fig. 8b, the contribution from stars at
$z > 2$ to 
the present ionizing radiation background just above the Lyman 
edge is less than 3\%.  This assumes that the emissivity
peaked at $z \approx2$ and falls off at high redshift.
If, as suggested by Pettini et al. (1998) and Steidel et al.
(1999), there is little or no falloff in the star formation rate out
to $z \approx 4$, the mean intensity at $z=0$ would increase
by only $10\%$.  This is a small effect because of redshifting
and the fact that stars emit few photons more 
energetic than 4 rydbergs.  Thus, stars at $z > 3$ would
make little contribution to $I_0$ at $z=0$.  Also, because stellar
radiation does not doubly ionize He, diffuse He~II Ly$\alpha$
and two-photon emission from absorbers make little contribution to
$I_0$ at $z=0$.

\section*{4. CONCLUSIONS }

In this paper, we have endeavored to make accurate estimates
of the low-redshift intensity of ionizing radiation, arising
from QSOs, Seyfert galaxies, and starburst galaxies.  In performing
this calculation, we found that we require accurate values 
of AGN emissivity and IGM opacity out to substantial redshifts.
In other words, this is a global problem.  

Our new estimates of the ionizing emissivities of Seyfert galaxies 
and low-redshift QSOs were constructed by extrapolating ultraviolet 
fluxes from IUE to the Lyman limit.  For starburst galaxies, we used 
recent H$\alpha$ surveys together with an educated guess for the
escaping fraction of ionizing radiation.  The IGM opacity was 
derived from HST surveys of the low-redshift Ly$\alpha$
absorbers and a new opacity model from {\it Keck} high-resolution
spectra of high-redshift QSOs.  By incorporating these
ingredients into a cosmological radiative transfer code, we 
find that the ionizing intensity at $z \approx 0$ has approximately
equal contributions from AGN and starburst galaxies: 
\begin{eqnarray}
   I_0^{\rm AGN} &=& (1.3 \times 10^{-23}~{\rm erg~cm}^{-2}~{\rm s}^{-1}~
       {\rm Hz}^{-1}~{\rm sr}^{-1})     
      \left[ \frac {L_{\rm min}}{0.01 L_*} \right] ^{-0.17} 
      \left[ \frac {\alpha_s}{1.8} \right] ^{-0.97}  \\
                       \,               \nonumber \\
I_0^{\rm Star} &=& (1.1 \times 10^{-23}~{\rm erg~cm}^{-2}~{\rm s}^{-1}~
       {\rm Hz}^{-1}~{\rm sr}^{-1})
      \left[ \frac {\langle f_{\rm esc} \rangle}{0.05} \right] 
      \left[ \frac {L_{H\alpha}} {10^{39.1}} \right] 
\end{eqnarray}
Taking into account uncertainties in the various parameters of
the model ($L_{\rm min}$, $\alpha_s$, QSO luminosity 
function, H$\alpha$-determined star-formation history) we 
estimate uncertainties in these coefficients of $1.3^{+0.8}_{-0.5}$ 
(for AGN) and $1.1^{+1.4}_{-0.7}$ (for starbursts). 
For a spectral index $\alpha_s \approx 1.8$ (from 1 -- 4 Ryd), 
these values of $I_0$ each correspond to one-sided ionizing fluxes
$\Phi_{\rm ion} \approx 3000$ photons cm$^{-2}$ s$^{-1}$.
Allowing for statistical uncertainties,
the total ionizing photon flux at low redshift 
probably lies in the range $\Phi_{\rm ion} = 2000 - 10,000$ 
photons cm$^{-2}$ s$^{-1}$, which is consistent with a number
of recent estimates and measurements (see Table 1). 

The redshift evolution of the hydrogen photoionization rate,
$\Gamma_{\rm HI}(z)$ is shown in Figure 9 for three cases:
AGN only, starbursts only, and combined (AGN plus starbursts).  
These rates were computed using 
our standard emissivity models for AGN and starburst galaxies,
except that the starburst emissivities were held constant
at $z > 1.7$ to simulate recent measurements at high $z$.  The 
starburst emissivities also assume $\langle f_{\rm esc} \rangle = 0.05$.   
Although it is beyond the scope of this paper, the potentially
dominant role of starburst galaxies in photoionization at $z > 4$ 
is apparent.

Because the values in eqs. (14) and (15) include unacceptably 
large range for such an important
quantity, it is worth discussing what might be done to improve
the situation, both thoretically and observationally.  
Advances need to be made in the characterization of both 
ionizing emissivities and IGM opacities.  
The greatest uncertainty in the opacity model occurs in the
range $\log N\subH = 15-17$, where line saturation and
small-number statistics make the Ly$\alpha$ surveys inaccurate.
The imminent launch of the FUSE satellite 
will open up the far-UV band (920--1180~\AA) that contains
Ly$\beta$ and higher Lyman series lines.  A survey of Ly$\beta$
lines should allow better determinations of line saturation
(doppler $b$-values).  The FUSE spectra of AGN will also
provide more accurate values of the flux near the Lyman limit
and reduce the uncertainties introduced by extrapolating from
spectral regions longward of 1200 \AA, as we have done with IUE data.

To address the general problem of small-number statistics
in the Ly$\alpha$ absorbers, the HST Cosmic Origins Spectrograph
(Morse et al. 1998), scheduled for installation on HST in 2003, 
should be used for a QSO absorption-line key project.  A full 
discussion of the advantages of this project is given in Appendix 1 
of the UV-Optical Working Group White Paper (Shull et al. 1999).  
Because COS will have about 20 times the far-UV throughput of 
the previous HST spectrographs, GHRS and STIS, and offers 
velocity resolution 10 times better than that of FOS, it will 
provide much better statistics on  the distribution of H~I 
absorbers in both space and column density. 
The most useful COS surveys will be of low-redshift 
Ly$\alpha$ lines, particularly the rare ``partial Lyman-limit 
systems'' ($16.0 < \log N\subH < 17.5$).  Accurately characterizing
the distribution in column density and the evolution in redshift of these
absorbers would remove a large part of the uncertainty in the IGM 
opacity model.   

The emissivity models for both AGN and starburst galaxies also need
improvement.  Although we have used current surveys of Seyfert
galaxies and QSOs, we may have
missed certain classes of sources that are strong emitters in the
EUV. We believe that BL Lac objects contribute
less than 10\% of Seyferts, based on estimates of their luminosity
function and space density.  On the other hand,  Edelson et al. (1999) 
suggest that narrow-line Seyfert 1 galaxies may account for $\sim50$\%
of the EUV volume emissivity in the ROSAT Wide-Field Camera sample.  
It is not clear whether these Seyferts are captured in the
Cheng \ea (1985) luminosity function, but their ionizing spectra might
be higher than that derived from an extrapolation of their UV fluxes.
For low-redshift starbursts, two recent surveys (Gallego et al. 1995; 
Gronwall 1998) derive comparable values of H$\alpha$ luminosity density, 
although even the latter (KISS) H$\alpha$ survey may still be incomplete 
at the faint end. At higher redshifts, the QSO luminosity   
function is uncertain, owing to the effects of dust (Fall \& Pei 1993;
Pei 1995) and faint-end survey incompleteness.  
QSO surveys by GALEX (Galaxy Explorer) in the ultraviolet 
and by the Sloan Digital Sky Survey in the optical may clarify 
the AGN luminosity functions.  However, it is worth repeating 
that, owing to redshifting and IGM opacity, 
AGN and starbursts at $z > 3$ contribute  less than 10\% 
to the value of $I_0$ at $z = 0$.  

For an accurate emissivity density, what is needed most are surveys
at $z < 1$ of AGN and starburst galaxies.  Even after we 
ascertain the space density and H$\alpha$ luminosity functions of 
star-forming galaxies, we still need an accurate measurement of 
$\langle f_{\rm esc} \rangle$, the fraction of stellar ionizing
photons that escape the galactic H~I layers into the halo and IGM.
Here, we have relied on recent theoretical work (Dove et al. 1999) and
H$\alpha$ observations of gas in the Magellanic Stream
(Bland-Hawthorn \& Maloney 1999) that suggest $\langle f_{\rm esc}
\rangle \approx 0.03-0.06$.  However, access with FUSE to the far-UV 
spectrum at 920--950 \AA\ allows a direct measurement of
the escaping EUV continuum from starbursts at redshifts
$z \approx 0.05$.  This work will extend the studies  
with the Hopkins Ultraviolet Telescope (HUT) 
of leaky starbursts (Leitherer et al. 1995; Hurwitz et al. 1997).

Finally, we eagerly await new measurements of the ionizing
photon flux, $\Phi_{\rm ion}$ via several direct and indirect
techniques.  These methods include improved
Fabry-Perot measurements of H$\alpha$ from Galactic 
high-velocity clouds (Tufte et al. 1998; Bland-Hawthorn \& Maloney 1999),
and UV absorption-line measurements of ionization
ratios such as Fe~I/Fe~II and Mg~I/Mg~II that constrain the
far-UV radiation in the 0.6--1.0 Ryd band (Stocke et al. 1991;
Tumlinson  et al. 1999). A new absorption-line key project 
with HST/COS could make precise estimates of $I_0$ from 
the proximity effect.   
As a result of the new surveys and ventures mentioned above,
it should be possible, within five years, to determine the
local ionizing background to $<30$\%.  With good
fortune, these measurements and the theoretical models will 
agree to a level better than described here in Table 1.

\vspace{1cm}

This work was supported by theoretical grants from NASA
(NAG5-7262) and NSF (AST96-17073).  The IUE spectral analysis 
was supported by a grant from NASA's Astrophysical Data Program
(NAG5-3006).  

  \newpage       

\begin{center}
{\bf Table 1} \\
{\bf Measurements and Limits of low-$z$ Ionizing Background$^1$ } \\
\   \\
\begin{tabular}{lrl}
\hline\hline 
Technique  &  $\Phi_{\rm ion}$ (cm$^{-2}$~s$^{-1}$)      &  Reference   \\   
\hline
H$\alpha$ Fabry-Perot        & $<3  \times 10^4$       & Vogel \ea (1995)   \\
H$\alpha$ Filter Images      & $<1.1 \times 10^4$      & Donahue \ea (1995)   \\
H$\alpha$ Filter Images      & $<8.4  \times 10^4$     & Stocke \ea (1991)   \\
H$\alpha$ Filter Images      & $<9 \times 10^4$        & Kutyrev \& Reynolds 
                                                           (1989) \\
H~I Disk Edges               & $(0.5-5) \times 10^4$   & Maloney (1993) \\ 
H~I Disk Edges               & $(1-5) \times 10^4$     & Dove \& Shull (1994a)   \\
Prox. Eff. $\langle z \rangle = 0.5$ & $(0.05-1.0) \times 10^4$ 
                                               & Kulkarni \& Fall (1993) \\ 
\hline \\
\end{tabular}
\end{center}

\noindent
$^1$ $\Phi_{\rm ion}$ is the one-sided, normally incident photon flux 
in the metagalactic
background, related to the specific intensity at Lyman limit by
$\Phi_{\rm ion} = (2630~{\rm cm}^{-2}~{\rm s}^{-1}) I_{-23} (1.8/\alpha_s)$ 
-- see  eq. (1) in text. 


\vspace{1cm}

\begin{center}
{\bf Table 2} \\
{\bf  Low-$z$ Opacity Models } \\
\   \\
\begin{tabular}{lrrrrrrl}
\hline\hline
Model &  $N_{\rm min}$(cm$^{-2}$) & $N_{\rm max}$(cm$^{-2}$) & $A_i$ & $\beta_i$
 & $\gamma_i$ & $c_i$ & $s_i$   \\
\hline
\\
Standard  & ...       & $10^{14}$  & 0.105 & 1.63 & 0.15 & 0.010 & -2.81
\\
          & $10^{14}$ & $10^{17}$  & 0.501 & 1.50 & 0.15 & 0.553 & -2.73
\\
          & $10^{17}$ & $10^{22}$  & 0.159 & 1.50 & 1.5  & 0.263 & -1.04
\\
Model 1   & ...       & $10^{15.5}$& 0.105 & 1.63 & 0.15 & 0.048 & -2.81
\\
          &$10^{15.5}$& $10^{22}$  & 0.159 & 1.50 & 1.5  & 0.411 & -1.38
\\
Model 2   & ...       & $10^{15.5}$& 0.105 & 1.63 & 0.15 & 0.048 & -2.81
\\
          & $10^{17}$ & $10^{22}$  & 0.159 & 1.50 & 1.5  & 0.263 & -1.04
\\
\\
\hline \\
\end{tabular}
\end{center}


\begin{figure}
\plotone{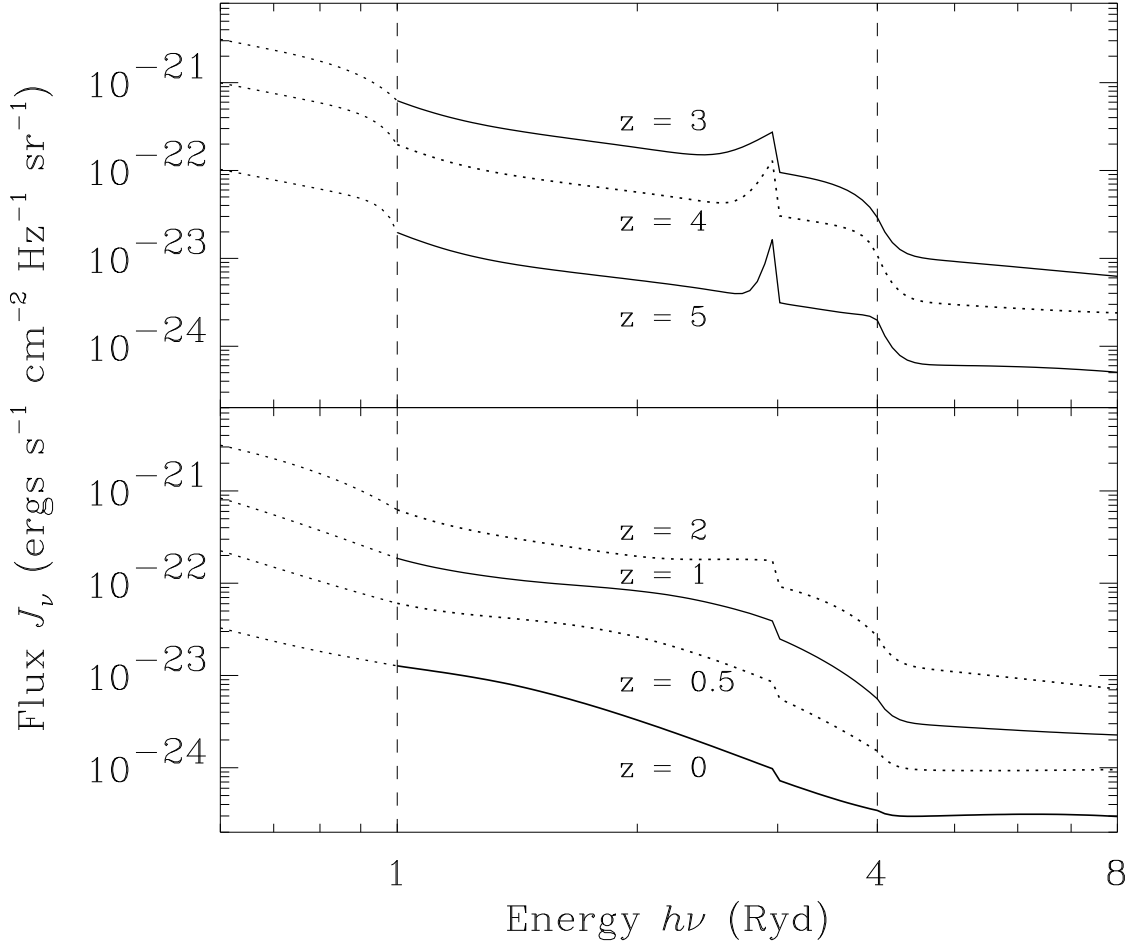}
\epsscale{0.8}
\caption{
Evolution of the ionizing background with redshift, assuming the 
modified Pei luminosity function for AGN described in \S~2.1 
($\Omega_0 = 0.2, h=0.5, \alpha_{\rm UV}=0.86, \alpha_s=1.8$).  The opacity 
is a hybrid model consisting of Model A2 from FGS for $z > 1.9$, and 
our standard model for the low redshift opacity for $z < 1.9$.
This calculation incorporates the full radiative transfer and 
reradiation of ionizing photons by absorbers described in FGS.
Top panel: $3 \le z \le 5$.  Bottom panel: $0 \le z \le 2$.  }
\end{figure}
 
\begin{figure}
\plotone{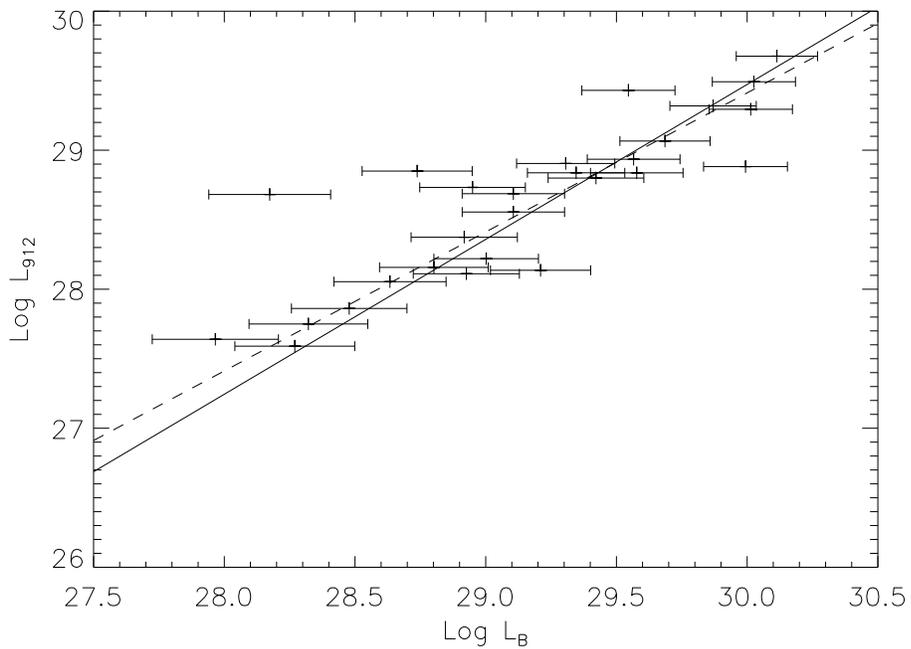}
\caption{Correlation between $L_{912}$ and $L_B$ for the 27 Seyferts
in our subsample.  A least-squares fit gives a slope of $1.114 \pm
0.081$ with $\chi^2 = 65.5$.  For comparison, the dashed curve shows
a constant optical spectral index of $\alpha_{\rm UV} = 0.86$ between
$4400$ \AA\ and $912$ \AA. } 
\end{figure}

\begin{figure}
\plotone{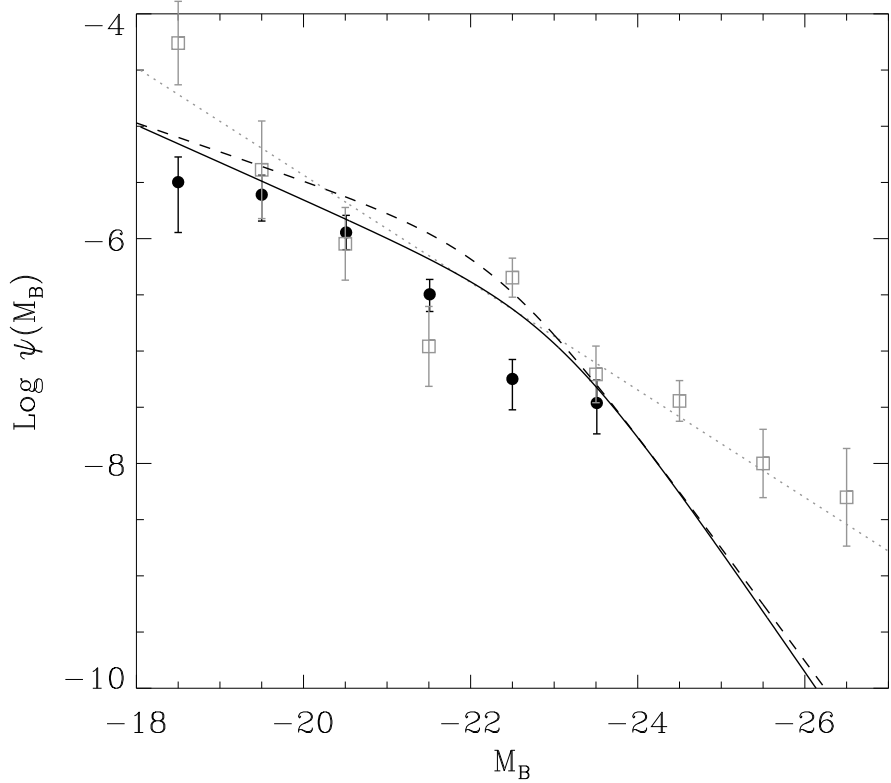}
\caption{ AGN luminosity functions.
Circles show the Seyfert galaxy sample of Cheng \ea (1995)
from which the 27 Seyfert galaxies in common
with our IUE-AGN database (Penton \ea 1998) were taken (see Fig. 2).
Squares show the Seyfert galaxy sample of K\"{o}hler \ea (1997),
and dotted curve represents a power law fit $\Phi(L_B) \propto L_B^{-2.2}$
to these latter results.  The solid and dashed curves are an extrapolation 
to $z=0$ of the Pei (1995) luminosity function for $h=0.5, \Omega_0=0.2$
and $h=0.5, \Omega_0 =1$, respectively.  Note that at $z = 0$, the AGN 
fitted by Pei only went down to $0.7L_\ast$ corresponding to $M_B = -23.5$.}
\end{figure}

\begin{figure}
\plotone{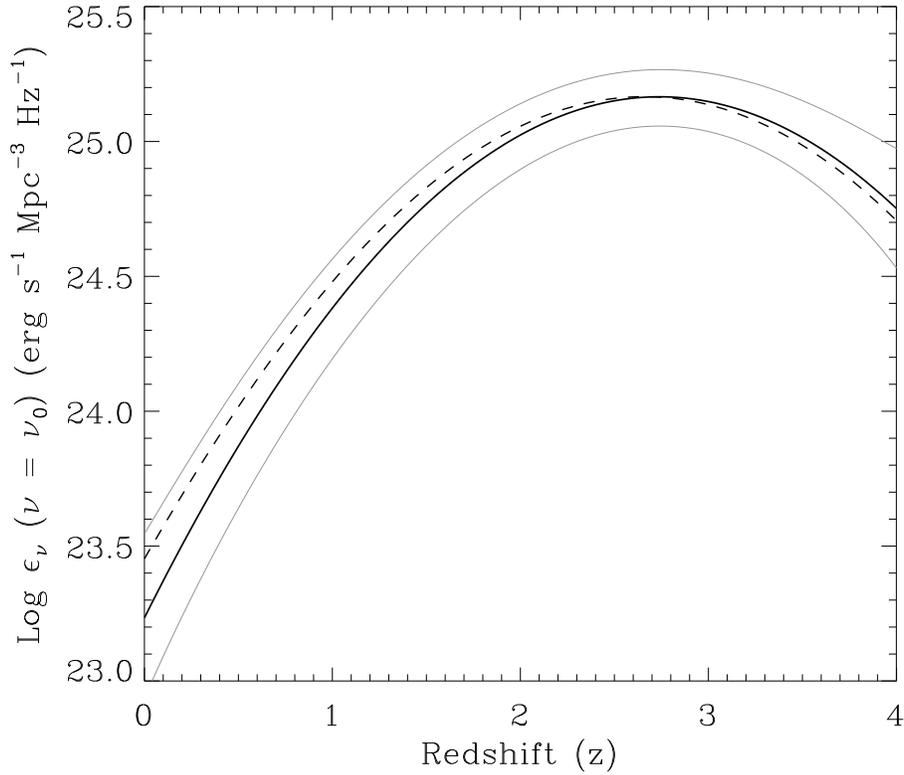}
\caption{Source emissivity due to AGN at redshift $z$ for
modified Pei open model ($\Omega_0=0.2$, $h=0.5$, and $\alpha_{\rm UV} = 0.86$) 
(solid curve) and modified Pei closed model ($\Omega_0=1$, $h=0.5$, and 
$\alpha_{\rm UV} = 0.86$; correction for open universe line element)
(dashed curve).  Dotted curves frame our $1 \sigma$ range in emissivity 
for the open model, including corrections for completeness of the 
luminosity function, and a variation of the $L_{912}/L_B$ ratio by 
one standard deviation in spectral slope and dependence on $L_B$.  }
\end{figure}

\begin{figure}
\plotone{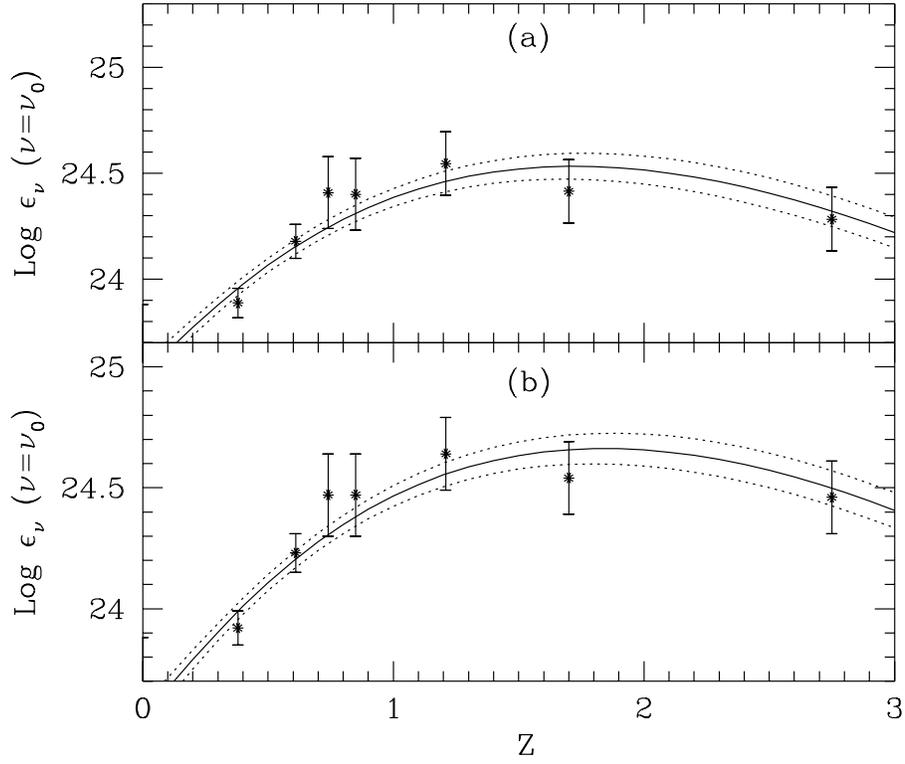}
\caption{(a)  Source emissivity due to galaxies at redshift $z$,
tied to Gallego \ea (1995) and Connolly \ea (1997).
Assumes $\Omega_0 = 0.2, h=0.5,$ and $\langle f_{esc} \rangle  = 0.05$,
$T= 2\times 10^4$ K, $\langle h\nu \rangle = 22$ eV.
Dotted curves frame our 1$\sigma$ range in emissivity.
(b)  Same as (a), except $\Omega_0 = 1$}
\end{figure}

\begin{figure}
\plotone{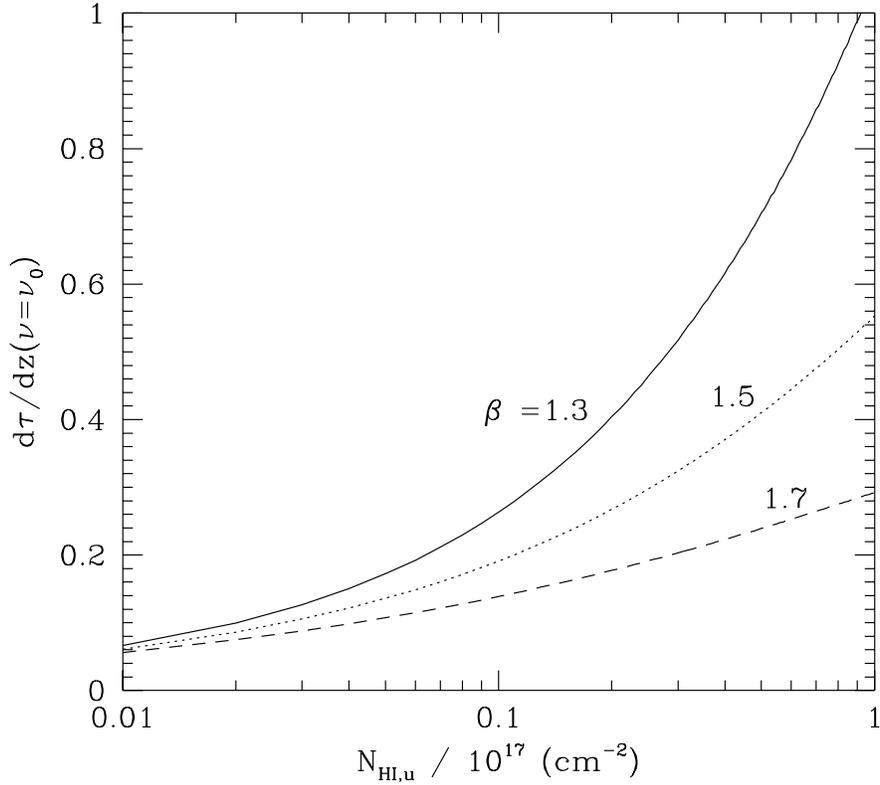}
\caption{Differential continuum opacity, $d \tau_{\rm eff}/dz$, at 
$\nu = \nu_0$ and $z=0$, versus the assumed upper limit on the column 
density distribution.  We consider HST/FOS sample 5 of Weymann \ea
(1998) of 465 Ly$\alpha$ absorbers with 
$W_{\lambda} > 240$~m\AA, for which $d{\cal N}/dz = 30.7 \pm 4.2$.
We model these lines as a distribution, $N_{HI}^{-\beta}$, 
of column densities from  $N_{HI} = 10^{14}$ cm$^{-2}$ up to $N_{HI,u}$.
In ascending order, these curves assume $\beta$ = 1.7, 1.5, and 1.3. } 
\end{figure}

\begin{figure}
\plotone{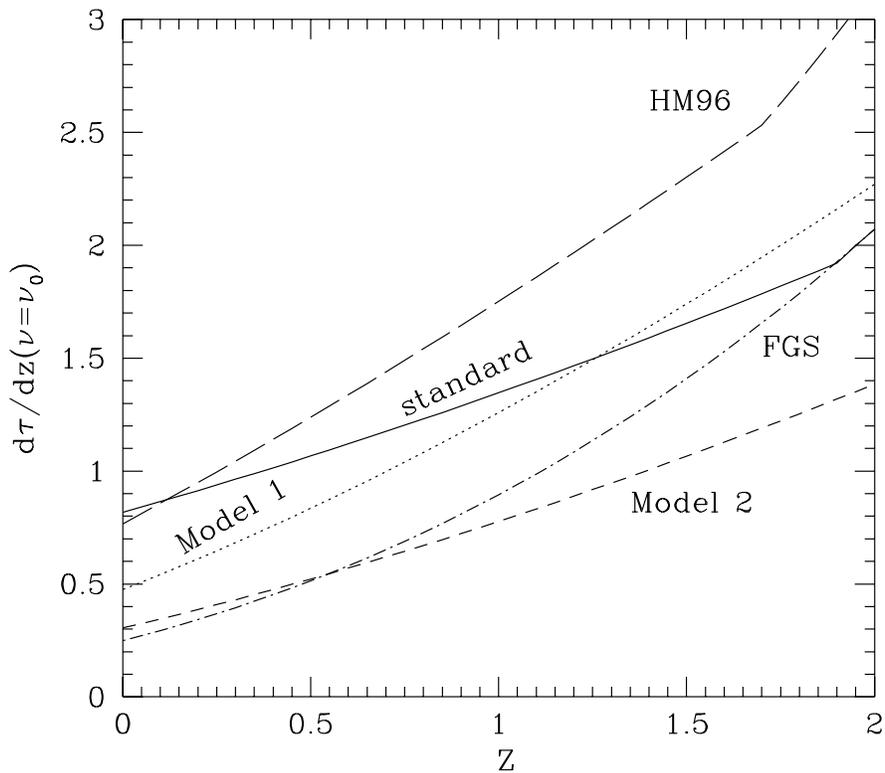}
\caption{Differential continuum opacity, 
$d \tau_{\rm eff}/dz$, at $\nu = \nu_0$ versus redshift
$z$ for models for the column density distribution of
Ly$\alpha$ forest and Lyman limit systems.  Solid curve
is our ``standard model'' described in Table 2.
Dotted curve is our Model 1, and short-dashed curve is our
model 2.  Dot-dashed curve is the extrapolation to low $z$ of
Model A2 from FGS.  Long-dashed curve is the opacity assumed by 
Haardt \& Madau (1996).  }
\end{figure}

\begin{figure}
\plotone{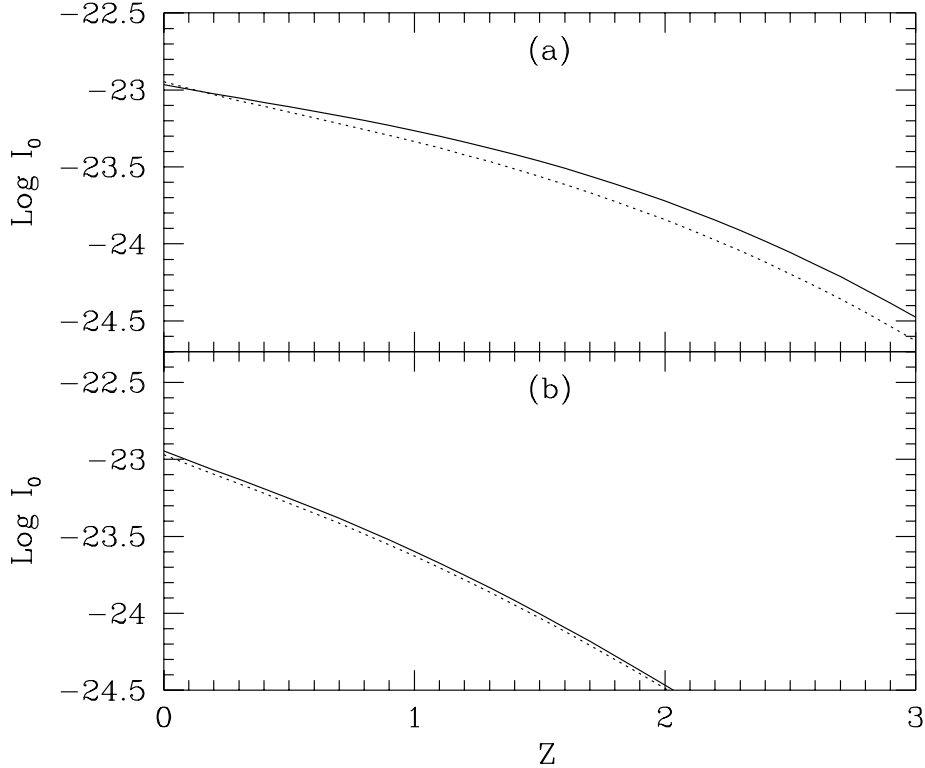}
\caption{(a) Cumulative value of $I_0(z=0)$ arising from AGN at 
redshifts greater than $z$.  These intensties are calculated using 
the approximate method (eqs. (3-5]) that neglects diffuse
ionizing radiation from the absorbers.  
Solid curve assumes $\Omega_0 = 0.2, h=0.5$; dotted curve assumes 
$\Omega_0 = 1, h=0.5$.  Standard model for opacity is used for both. 
At $z=0$, $I_0 = 1.09 \times 10^{-23}$ and $I_0 = 1.13 \times 10^{-23}$ 
for these respective cases. 
(b)  Cumulative value of $I_0$ arising from galaxies at redshift $z$,
assuming $\Omega_0 = 0.2, h=0.5,$ (solid curve) or
$\Omega_0 = 1, h=0.5,$ (dotted curve), $\langle f_{esc} \rangle = 0.05$,
and standard model for opacity.  At $z = 0$,  
$I_0 = 1.13 \times 10^{-23}$ and $I_0 = 1.07 \times 10^{-23}$ 
for the respective cases. }
\end{figure}

\begin{figure}
\plotone{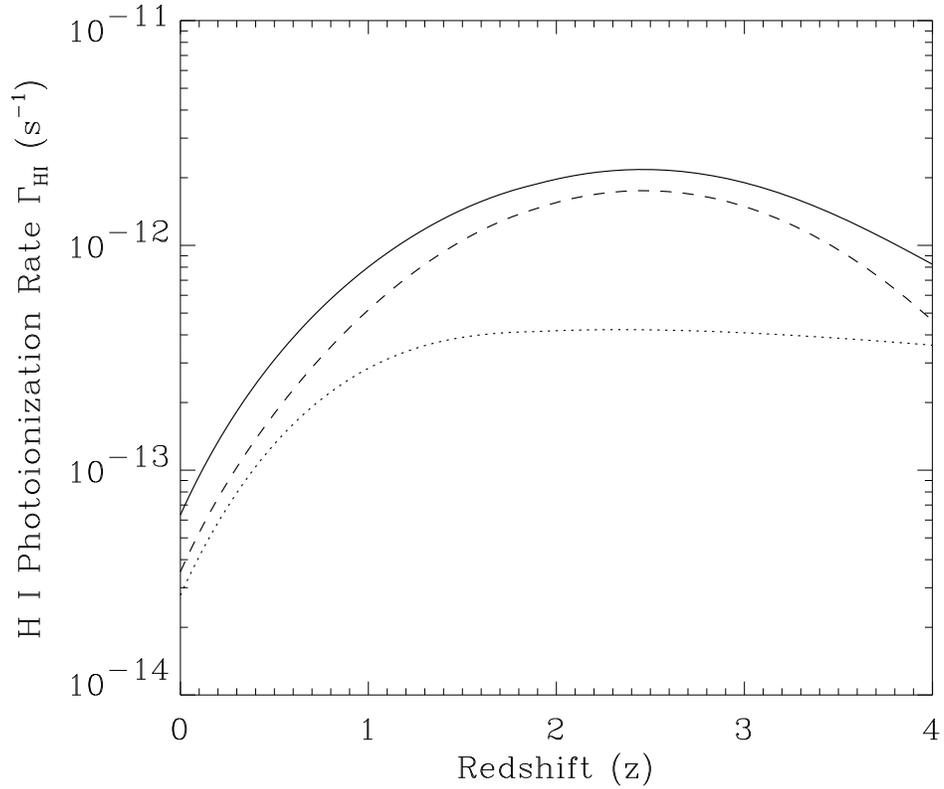}
\caption{Hydrogen photoionization rate versus redshift for three cases:
AGN-only (dashed curve), starburst galaxies only (dotted), and combined 
(solid curve). Each case assumes the standard models for emissivity
(\S~2.1, 2.2) and opacity (\S2.3 and Fig. 1 caption).  The AGN emissivities
are the same as solid curve in Fig. 4, while starburst emissivities are
same as solid curve in Fig. 5a, except that the comoving emissivity is held constant
for $1.7 < z < 4.0$ to simulate recent measurements of star-formation history.} 
\end{figure}
\end{document}